\documentstyle[12pt]{article}
\begin{document}

\author{C. Bizdadea\thanks{%
e-mail addresses: bizdadea@central.ucv.ro and bizdadea@hotmail.com}, E. M.
Cioroianu\thanks{%
e-mail address: manache@central.ucv.ro}, S. O. Saliu\thanks{%
e-mail addresses: osaliu@central.ucv.ro and odile\_saliu@hotmail.com} \\
Department of Physics, University of Craiova\\
13 A. I. Cuza Str., Craiova RO-1100, Romania}
\title{Irreducible Hamiltonian BRST symmetry for 
reducible first-class systems }
\maketitle

\begin{abstract}
An irreducible Hamiltonian BRST quantization method for reducible
first-class systems is proposed. The general theory is illustrated on a
two-stage reducible model, the link with the standard reducible BRST
treatment being also emphasized.

\end{abstract}

\section{Introduction}

It is well known that the Hamiltonian BRST formalism \cite{1}--\cite{6}
stands for one of the strongest and most popular quantization methods for
theories with first-class constraints. This method can be applied to
irreducible, as well as reducible first-class systems. In the irreducible
case the ghosts can be regarded as one-forms dual to the vector fields
associated with the first-class constraints. This interpretation fails in
the reducible framework, being necessary to introduce ghosts for ghosts
together with their canonical conjugated momenta (antighosts). The ghosts
for ghosts ensure the incorporation of the reducibility relations within the
cohomology of the exterior derivative along the gauge orbits, while their
canonical momenta are required in order to kill the higher-resolution-degree
nontrivial co-cycles from the homology of the Koszul-Tate differential.
Similar considerations apply to the antifield BRST treatment \cite{6}--\cite
{11}. 

Another way of approaching reducible systems is based on the idea of
replacing such a system with an irreducible one \cite{6}, \cite{12}. This
idea has been enforced in the framework of the irreducible quantization of
reducible theories from the point of view of the antifield-BRST formulation 
\cite{ir1}, as well as of the antifield BRST-anti-BRST method \cite{ir2}.
Some applications of these treatments can be found in \cite{ir3}--\cite{ir4}%
. At the level of the Hamiltonian symmetry, this type of procedure has been
developed so far only in the case of some peculiar models \cite{ir5}--\cite
{ir6}, but a general theory covering on-shell reducible first-class systems
has not yet been given. 

In view of this, here we propose an irreducible Hamiltonian BRST procedure
for quantizing on-shell reducible first-class theories. In this light, we
enforce the following steps: (i) we transform the original reducible
first-class constraints into some irreducible ones on a larger phase-space
in a manner that allows the substitution of the BRST quantization of the
reducible system by that of the irreducible theory; (ii) we quantize the
irreducible system accordingly the Hamiltonian BRST formalism. As a
consequence, the ghosts for ghosts and their antighosts do not appear in our
formalism. 

The paper is organized in five sections. Section 2 deals with the derivation
of an irreducible set of first-class constraints associated with the
original reducible one by means of constructing an irreducible Koszul-Tate
complex. The irreducible Koszul-Tate complex is obtained by requiring that
all the antighost number one co-cycles from the reducible case become
trivial under an appropriate redefinition of the antighost number one
antighosts. This procedure leads to the introduction of new canonical
variables and antighosts. In section 3 we infer the irreducible BRST
symmetry corresponding to the irreducible constraint set deduced in section
2 and prove that we can replace the Hamiltonian BRST quantization of the
reducible system with that of the irreducible theory. Section 4 illustrates
our method in the case of a two-stage reducible model involving three-form
gauge fields. In section 5 we present the main conclusions of the paper.

\section{Derivation of an irreducible first-class constraint set.
Irreducible Koszul-Tate differential}

\subsection{Setting the problem}

Our starting point is a Hamiltonian system with the phase-space locally
described by $N$ canonical pairs $z^{A}=\left( q^{i},p_{i}\right) $, subject
to the first-class constraints 
\begin{equation}
G_{a_{0}}\left( q^{i},p_{i}\right) \approx 0,\;a_{0}=1,\cdots ,M_{0},
\label{1}
\end{equation}
which are assumed to be on-shell $L$-stage reducible. We suppose that there
are no second-class constraints in the theory (if any, they can be
eliminated with the help of the Dirac bracket). The first-class property of
the constraints (\ref{1}) is expressed by 
\begin{equation}
\left[ G_{a_{0}},G_{b_{0}}\right] =C_{\;\;a_{0}b_{0}}^{c_{0}}G_{c_{0}},
\label{2}
\end{equation}
while the reducibility relations are written as 
\begin{equation}
Z_{\;\;a_{1}}^{a_{0}}G_{a_{0}}=0,\;a_{1}=1,\cdots ,M_{1},  \label{3}
\end{equation}
\begin{equation}
Z_{\;\;a_{2}}^{a_{1}}Z_{\;\;a_{1}}^{a_{0}}=C_{a_{2}}^{a_{0}b_{0}}G_{b_{0}},%
\;a_{2}=1,\cdots ,M_{2},  \label{4}
\end{equation}
\[
\vdots 
\]
\begin{equation}
Z_{\;\;a_{L}}^{a_{L-1}}Z_{\;%
\;a_{L-1}}^{a_{L-2}}=C_{a_{L}}^{a_{L-2}b_{0}}G_{b_{0}},\;a_{L}=1,\cdots
,M_{L},  \label{5}
\end{equation}
with the symbol $\left[ ,\right] $ denoting the Poisson bracket. For
definiteness we approach here the bosonic case, but our analysis can be
easily extended to fermions modulo introducing some appropriate sign
factors. The first-order structure functions $C_{\;\;a_{0}b_{0}}^{c_{0}}$
may involve the phase-space coordinates (open gauge algebra) and are
antisymmetric in the lower indices. The reducibility functions $\left(
Z_{\;\;a_{k+1}}^{a_{k}}\right) _{k=0,\cdots ,L-1}$ and the coefficients $%
\left( C_{a_{k+2}}^{a_{k}b_{0}}\right) _{k=0,\cdots ,L-2}$ appearing in the
right hand-side of the relations (\ref{4}--\ref{5}) may also depend on the
canonical variables, and, in addition, $C_{a_{2}}^{a_{0}b_{0}}$ should be
antisymmetric in the upper indices.

The standard Hamiltonian BRST symmetry for the above on-shell reducible
first-class Hamiltonian system, $s_R=\delta _R+D_R+\cdots $, contains two
crucial operators. The Koszul-Tate differential $\delta _R$ realizes an
homological resolution of smooth functions defined on the first-class
constraint surface (\ref{1}), while the model of longitudinal exterior
derivative along the gauge orbits $D_R$ is a differential modulo $\delta _R$
and accounts for the gauge invariances. The degree of $\delta _R$ is called
antighost number ($antigh$), the degree of $D_R$ is named pure ghost number (%
$pgh$), while the overall degree of the BRST differential is called ghost
number ($gh$) and is defined like the difference between the pure ghost
number and the antighost number ($antigh\left( \delta _R\right) =-1$, $%
pgh\left( D_R\right) =1$, $gh\left( s_R\right) =1$). In the case of a
first-stage reducible Hamiltonian system ($L=1$) the proper construction of $%
\delta _R$ relies on the introduction of the generators (antighosts) ${\cal P%
}_{a_0}$ and $P_{a_1}$, with the Grassmann parity ($\epsilon $) and the
antighost number given by 
\begin{equation}  \label{6}
\epsilon \left( {\cal P}_{a_0}\right) =1,\;\epsilon \left( P_{a_1}\right)
=0,\;antigh\left( {\cal P}_{a_0}\right) =1,\;antigh\left( P_{a_1}\right) =2,
\end{equation}
on which $\delta _R$ is set to act like 
\begin{equation}  \label{7}
\delta _R{\cal P}_{a_0}=-G_{a_0},
\end{equation}
\begin{equation}  \label{8}
\delta _RP_{a_1}=-Z_{\;\;a_1}^{a_0}{\cal P}_{a_0},
\end{equation}
being understood that 
\begin{equation}  \label{9}
\delta _Rz^A=0.
\end{equation}
The antighosts $P_{a_1}$ are required in order to ``kill'' the antighost
number one nontrivial co-cycles 
\begin{equation}  \label{10}
\mu _{a_1}=Z_{\;\;a_1}^{a_0}{\cal P}_{a_0},
\end{equation}
(which are due to the definitions (\ref{7}) and the reducibility relations (%
\ref{3})) in the homology of $\delta _R$, and thus restore the acyclicity of
the Koszul-Tate differential at nonvanishing antighost numbers. For a
two-stage reducible Hamiltonian system ($L=2$), one should supplement the
antighost spectrum from the first-stage case with the antighosts $P_{a_2}$,
displaying the characteristics $\epsilon \left( P_{a_2}\right) =1$, $%
antigh\left( P_{a_2}\right) =3$, and define the action of $\delta _R$ on
them through $\delta _RP_{a_2}=-Z_{\;\;a_2}^{a_1}P_{a_1}-\frac
12C_{a_2}^{a_0b_0}{\cal P}_{b_0}{\cal P}_{a_0}$. The acyclicity of $\delta
_R $ is thus achieved by making exact the antighost number two nontrivial
co-cycles $\mu _{a_2}=Z_{\;\;a_2}^{a_1}P_{a_1}+\frac 12C_{a_2}^{a_0b_0}{\cal %
P}_{b_0}{\cal P}_{a_0}$, which are present on behalf of the definitions (\ref
{8}) and the reducibility relations (\ref{4}). The process goes along the
same lines at higher antighost numbers. In the general situation of an $L$%
-stage reducible Hamiltonian theory, the antighost spectrum contains the
variables ${\cal P}_{a_0}$ and $\left( P_{a_k}\right) _{k=1,\cdots ,L}$,
with $\epsilon \left( P_{a_k}\right) =k+1\;mod\;2$, $antigh\left(
P_{a_k}\right) =k+1$, the actions of $\delta _R$ on higher antighost number
antighosts being taken in such a way to ensure the acyclicity of the
Koszul-Tate differential at nonvanishing antighost numbers.

The problem to be investigated in the sequel is the derivation of an
irreducible set of first-class constraints associated with the $L$-stage
reducible one (\ref{1}). Our basic idea is to redefine the antighosts ${\cal %
P}_{a_{0}}$ in such a way that the new co-cycles of the type (\ref{10}) are
trivial. Then, the antighosts $P_{a_{1}}$ ensuring the triviality of the
co-cycles (\ref{10}) are no longer necessary, so they will be discarded from
the Koszul-Tate complex. Moreover, the absence of these antighosts implies
the absence of nontrivial co-cycles at antighost number greater that one,
hence the antighosts $\left( P_{a_{k}}\right) _{k=2,\cdots ,L}$ will also be
discarded from the antighost spectrum. The enforcement of this idea leads to
an irreducible set of first-class constraints underlying some physical
observables that coincide with those deriving from the original reducible
system. In order to simplify the presentation we initially approach the
cases $L=1,2$, and further generalize the results to an arbitrary $L$.

\subsection{The case $L=1$}

We begin with the definitions (\ref{7}--\ref{8}) and the reducibility
relations (\ref{3}). As we have previously mentioned, we redefine the
antighosts ${\cal P}_{a_{0}}$ like 
\begin{equation}
{\cal P}_{a_{0}}\rightarrow {\cal P}_{a_{0}}^{\prime }={\cal P}%
_{a_{0}}-Z_{\;\;b_{1}}^{b_{0}}\bar{D}_{\;\;c_{1}}^{b_{1}}A_{a_{0}}^{\;%
\;c_{1}}{\cal P}_{b_{0}},  \label{11}
\end{equation}
where $\bar{D}_{\;\;c_{1}}^{b_{1}}$ stands for the inverse of $%
D_{\;\;c_{1}}^{b_{1}}=Z_{\;\;c_{1}}^{c_{0}}A_{c_{0}}^{\;\;b_{1}}$, while $%
A_{a_{0}}^{\;\;c_{1}}$ are some functions that may involve at most the
variables $z^{A}$ and are chosen to satisfy $rank\left(
D_{\;\;c_{1}}^{b_{1}}\right) =M_{1}$. Next, we replace (\ref{7}) with 
\begin{equation}
\delta {\cal P}_{a_{0}}^{\prime }=-G_{a_{0}}.  \label{12}
\end{equation}
The definitions (\ref{12}) imply some co-cycles of the type (\ref{10}) 
\begin{equation}
\mu _{a_{1}}^{\prime }=Z_{\;\;a_{1}}^{a_{0}}{\cal P}_{a_{0}}^{\prime },
\label{13}
\end{equation}
which are found trivial on behalf of (\ref{11}), namely, $\mu
_{a_{1}}^{\prime }\equiv 0$. Thus, the definitions (\ref{12}) do not lead to
nontrivial co-cycles at antighost number one, therefore the antighosts $%
P_{a_{1}}$ will be discarded. Moreover, formulas (\ref{12}) are helpful at
deriving some irreducible first-class constraints corresponding to (\ref{1}%
). For this reason in (\ref{12}) we used the notation $\delta $ instead of $%
\delta _{R}$. The derivation of the irreducible first-class constraints
relies on enlarging the phase-space with the bosonic canonical pairs $\left(
y^{a_{1}},\pi _{a_{1}}\right) $, where the momenta $\pi _{a_{1}}$ are
demanded to be nonvanishing solutions to the equations 
\begin{equation}
D_{\;\;a_{1}}^{b_{1}}\pi _{b_{1}}=\delta \left( -Z_{\;\;a_{1}}^{a_{0}}{\cal P%
}_{a_{0}}\right) .  \label{14}
\end{equation}
As $D_{\;\;a_{1}}^{b_{1}}$ is invertible, the nonvanishing solutions to (\ref
{14}) implement the irreducibility. This is because the equations (\ref{14})
possess nonvanishing solutions if and only if 
\begin{equation}
\delta \left( Z_{\;\;a_{1}}^{a_{0}}{\cal P}_{a_{0}}\right) \neq 0,
\label{15}
\end{equation}
hence if and only if (\ref{10}) are not co-cycles. Inserting (\ref{14}) in (%
\ref{12}) and using (\ref{11}), we arrive at 
\begin{equation}
\delta {\cal P}_{a_{0}}=-G_{a_{0}}-A_{a_{0}}^{\;\;a_{1}}\pi _{a_{1}},
\label{16}
\end{equation}
which emphasize the irreducible constraints 
\begin{equation}
\gamma _{a_{0}}\equiv G_{a_{0}}+A_{a_{0}}^{\;\;a_{1}}\pi _{a_{1}}\approx 0.
\label{17}
\end{equation}
From (\ref{17}) it is easy to see that 
\begin{equation}
\pi _{a_{1}}=\bar{D}_{\;\;a_{1}}^{b_{1}}Z_{\;\;b_{1}}^{b_{0}}\gamma
_{b_{0}},\;G_{a_{0}}=\left( \delta _{\;\;a_{0}}^{b_{0}}-Z_{\;\;b_{1}}^{b_{0}}%
\bar{D}_{\;\;a_{1}}^{b_{1}}A_{a_{0}}^{\;\;a_{1}}\right) \gamma _{b_{0}},
\label{18}
\end{equation}
so 
\begin{equation}
\left[ \gamma _{a_{0}},\gamma _{b_{0}}\right] =\bar{C}_{\;%
\;a_{0}b_{0}}^{c_{0}}\gamma _{c_{0}},  \label{19}
\end{equation}
for some $\bar{C}_{\;\;a_{0}b_{0}}^{c_{0}}$. Thus, the irreducible
constraints (\ref{17}) are first-class. In the meantime, if we take the
standard action of $\delta $ on the phase-space coordinates 
\begin{equation}
\delta z^{A}=0,\;\delta z^{A_{1}}=0,  \label{20}
\end{equation}
with $z^{A_{1}}=\left( y^{a_{1}},\pi _{a_{1}}\right) $, then formulas (\ref
{16}) and (\ref{20}) completely define an irreducible Koszul-Tate complex
corresponding to an irreducible system based on the first-class constraints (%
\ref{17}).

At this point we underline two important observations. First, the number of
physical degrees of freedom of the irreducible system coincides with the
original one. Indeed, in the reducible case there are $N$ canonical pairs
and $M_{0}-M_{1}$ independent first-class constraints, hence $N-M_{0}+M_{1}$
physical degrees of freedom. In the irreducible situation there are $N+M_{1}$
canonical pairs and $M_{0}$ independent first-class constraints, therefore
as many physical degrees of freedom as in the reducible case. Second, from (%
\ref{14}) it results (due to the invertibility of $D_{\;\;a_{1}}^{b_{1}}$)
that the momenta $\pi _{a_{1}}$ are $\delta $-exact. These results represent
two desirable features, which will be requested also in connection with
higher-order reducible Hamiltonian systems. Anticipating a bit, we notice
that for higher-order reducible theories it is necessary to further add some
supplementary canonical variables and antighost number one antighosts. While
the former request indicates the number of new canonical pairs and
first-class constraints to be added within the irreducible framework, the
latter ensures, as it will be further seen, that there exists a proper
redefinition of the antighost number one antighosts that makes trivial the
co-cycles from the reducible approach.

\subsection{The case $L=2$}

Now, we start from the reducibility relations (\ref{3}--\ref{4}) and intend
to preserve the definitions (\ref{16}) and (\ref{20}) with respect to the
irreducible Koszul-Tate differential $\delta $. Nevertheless, there appear
two difficulties. First, the matrix $D_{\;\;c_{1}}^{b_{1}}$ is not
invertible now due to the second-stage reducibility relations (\ref{4}).
Rather, it possesses $Z_{\;\;a_{2}}^{c_{1}}$ as on-shell null vectors 
\begin{equation}
D_{\;\;c_{1}}^{b_{1}}Z_{\;\;a_{2}}^{c_{1}}=A_{c_{0}}^{\;\;b_{1}}Z_{\;%
\;c_{1}}^{c_{0}}Z_{\;\;a_{2}}^{c_{1}}=C_{a_{2}}^{c_{0}d_{0}}A_{c_{0}}^{\;%
\;b_{1}}G_{d_{0}}\approx 0,  \label{21}
\end{equation}
hence the transformations (\ref{11}) fail to be correct. We will still
maintain the definitions (\ref{16}) and will subsequently show that there
exists a redefinition of the antighosts ${\cal P}_{a_{0}}$ that brings the
constraint functions $\gamma _{a_{0}}$ under the form (\ref{17}). Second,
the irreducible first-class constraints (\ref{17}) are not enough in order
to maintain the initial number of physical degrees of freedom in connection
with the irreducible theory, so they should be supplemented with $M_{2}$ new
constraints in such a way to ensure a first-class irreducible behaviour of
the overall constraint set. Thus, we must introduce some fermionic antighost
number one antighosts ${\cal P}_{a_{2}}$ and set 
\begin{equation}
\delta {\cal P}_{a_{2}}=-\gamma _{a_{2}},  \label{22}
\end{equation}
where $\gamma _{a_{2}}\approx 0$ denote the new first-class constraints. On
the one hand, we should infer the concrete form of $\gamma _{a_{2}}$ such
that the new Koszul-Tate complex based on the definitions (\ref{16}), (\ref
{20}) and (\ref{22}) is irreducible, and, on the other hand, we should prove
that it is possible to perform a redefinition of the antighosts ${\cal P}%
_{a_{0}}$ (eventually involving also the new antighosts) such that (\ref{16}%
) is gained. This goes as follows. We begin with the relations (\ref{16}) on
which we apply $Z_{\;\;b_{1}}^{a_{0}}$, and obtain 
\begin{equation}
\delta \left( Z_{\;\;b_{1}}^{a_{0}}{\cal P}_{a_{0}}\right)
=-D_{\;\;b_{1}}^{a_{1}}\pi _{a_{1}}.  \label{23}
\end{equation}
Formulas (\ref{21}) make permissible a representation of $%
D_{\;\;b_{1}}^{a_{1}}$ under the form 
\begin{equation}
D_{\;\;b_{1}}^{a_{1}}=\delta _{\;\;b_{1}}^{a_{1}}-Z_{\;\;a_{2}}^{a_{1}}\bar{D%
}_{\;\;b_{2}}^{a_{2}}A_{b_{1}}^{\;\;b_{2}}+A_{a_{0}}^{\;%
\;a_{1}}C_{c_{2}}^{a_{0}b_{0}}\bar{D}_{\;\;b_{2}}^{c_{2}}A_{b_{1}}^{\;%
\;b_{2}}G_{b_{0}},  \label{24}
\end{equation}
where $\bar{D}_{\;\;b_{2}}^{a_{2}}$ is the inverse of $D_{\;%
\;b_{2}}^{a_{2}}=Z_{\;\;b_{2}}^{a_{1}}A_{a_{1}}^{\;\;a_{2}}$ and $%
A_{a_{1}}^{\;\;a_{2}}$ are some functions that may depend at most on $z^{A}$
and are taken to fulfill $rank\left( D_{\;\;b_{2}}^{a_{2}}\right) =M_{2}$.
Inserting (\ref{24}) in (\ref{23}), we infer that 
\begin{equation}
\delta \left( Z_{\;\;b_{1}}^{a_{0}}{\cal P}_{a_{0}}\right) =-\pi
_{b_{1}}+Z_{\;\;a_{2}}^{a_{1}}\bar{D}_{\;\;b_{2}}^{a_{2}}A_{b_{1}}^{\;%
\;b_{2}}\pi _{a_{1}}-C_{c_{2}}^{a_{0}b_{0}}\bar{D}_{\;%
\;b_{2}}^{c_{2}}A_{b_{1}}^{\;\;b_{2}}A_{a_{0}}^{\;\;a_{1}}\pi
_{a_{1}}G_{b_{0}},  \label{25}
\end{equation}
which turns into 
\begin{equation}
\delta \left( Z_{\;\;b_{1}}^{a_{0}}{\cal P}_{a_{0}}-C_{c_{2}}^{a_{0}b_{0}}%
\bar{D}_{\;\;b_{2}}^{c_{2}}A_{b_{1}}^{\;\;b_{2}}G_{b_{0}}{\cal P}%
_{a_{0}}\right) =-\pi _{b_{1}}+\bar{D}_{\;\;b_{2}}^{a_{2}}A_{b_{1}}^{\;%
\;b_{2}}Z_{\;\;a_{2}}^{a_{1}}\pi _{a_{1}},  \label{26}
\end{equation}
on account on the one hand of the antisymmetry of $C_{c_{2}}^{a_{0}b_{0}}$,
which implies the relations $C_{c_{2}}^{a_{0}b_{0}}A_{a_{0}}^{\;\;a_{1}}\pi
_{a_{1}}G_{b_{0}}=C_{c_{2}}^{a_{0}b_{0}}\gamma _{a_{0}}G_{b_{0}}$, and, on
the other hand, of the definitions (\ref{16}) and (\ref{20}). The second
term in the right hand-side of (\ref{26}) implies that $\pi _{a_{1}}$ are
not $\delta $-exact. An elegant manner of complying with the requirement on
the $\delta $-exactness of these momenta is to take the functions $\gamma
_{a_{2}}$ like 
\begin{equation}
\gamma _{a_{2}}\equiv Z_{\;\;a_{2}}^{a_{1}}\pi _{a_{1}},  \label{27}
\end{equation}
so $\delta {\cal P}_{a_{2}}=-Z_{\;\;a_{2}}^{a_{1}}\pi _{a_{1}}$, which then
leads, via (\ref{26}), to 
\begin{equation}
\pi _{b_{1}}=\delta \left( -Z_{\;\;b_{1}}^{a_{0}}{\cal P}%
_{a_{0}}+C_{c_{2}}^{a_{0}b_{0}}\bar{D}_{\;\;b_{2}}^{c_{2}}A_{b_{1}}^{\;%
\;b_{2}}G_{b_{0}}{\cal P}_{a_{0}}-\bar{D}_{\;\;b_{2}}^{a_{2}}A_{b_{1}}^{\;%
\;b_{2}}{\cal P}_{a_{2}}\right) .  \label{28}
\end{equation}
On behalf of (\ref{27}) and of the fact that the functions $%
Z_{\;\;a_{2}}^{a_{1}}$ have no null vectors, we find that (\ref{22}) imply
no nontrivial co-cycles. In this way the first task is achieved. Introducing
(\ref{28}) in (\ref{16}), we arrive at 
\begin{equation}
\delta {\cal P}_{a_{0}}^{\prime \prime }=-G_{a_{0}},  \label{29}
\end{equation}
where 
\begin{equation}
{\cal P}_{a_{0}}^{\prime \prime }={\cal P}_{a_{0}}-Z_{\;%
\;b_{1}}^{b_{0}}A_{a_{0}}^{\;\;b_{1}}{\cal P}_{b_{0}}+C_{c_{2}}^{c_{0}b_{0}}%
\bar{D}_{\;\;b_{2}}^{c_{2}}A_{b_{1}}^{\;\;b_{2}}A_{a_{0}}^{\;%
\;b_{1}}G_{b_{0}}{\cal P}_{c_{0}}-\bar{D}_{\;\;b_{2}}^{a_{2}}A_{b_{1}}^{\;%
\;b_{2}}A_{a_{0}}^{\;\;b_{1}}{\cal P}_{a_{2}}.  \label{30}
\end{equation}
Formula (\ref{30}) expresses a redefinition of the antighost number one
antighosts that is in agreement with (\ref{16}). It is precisely the
requirement on the $\delta $-exactness of the momenta $\pi _{a_{1}}$ that
allows us to deduce (\ref{30}). Thus, the second task is also attained.

It remains to be proved that (\ref{29}) also gives no nontrivial co-cycles.
From (\ref{29}) we obtain the co-cycles 
\begin{equation}
\mu _{a_{1}}^{\prime \prime }=Z_{\;\;a_{1}}^{a_{0}}{\cal P}_{a_{0}}^{\prime
\prime },  \label{31}
\end{equation}
at antighost number one. After some computation, we find that they are
trivial as 
\begin{equation}
\mu _{a_{1}}^{\prime \prime }=\delta \left( -\frac{1}{2}%
C_{c_{2}}^{a_{0}b_{0}}\bar{D}_{\;\;b_{2}}^{c_{2}}A_{a_{1}}^{\;\;b_{2}}{\cal P%
}_{a_{0}}^{\prime \prime }{\cal P}_{b_{0}}^{\prime \prime }\right) .
\label{32}
\end{equation}
In conclusion, we associated an irreducible Koszul-Tate complex based on the
definitions (\ref{16}), (\ref{20}) and (\ref{22}) with the starting
second-stage reducible one. This complex underlies the irreducible
constraint functions (\ref{17}) and (\ref{27}), which can be shown to be
first-class. Indeed, with the help of (\ref{17}) and (\ref{27}) we get that 
\begin{equation}
\pi _{b_{1}}=\left( Z_{\;\;b_{1}}^{a_{0}}-C_{c_{2}}^{a_{0}b_{0}}\bar{D}%
_{\;\;b_{2}}^{c_{2}}A_{b_{1}}^{\;\;b_{2}}G_{b_{0}}\right) \gamma _{a_{0}}+%
\bar{D}_{\;\;b_{2}}^{a_{2}}A_{b_{1}}^{\;\;b_{2}}\gamma _{a_{2}},  \label{34}
\end{equation}
\begin{equation}
G_{a_{0}}=\left( \delta
_{\;\;a_{0}}^{c_{0}}-Z_{\;\;b_{1}}^{c_{0}}A_{a_{0}}^{\;%
\;b_{1}}+C_{c_{2}}^{c_{0}b_{0}}\bar{D}_{\;\;b_{2}}^{c_{2}}A_{b_{1}}^{\;%
\;b_{2}}A_{a_{0}}^{\;\;b_{1}}G_{b_{0}}\right) \gamma _{c_{0}}-\bar{D}%
_{\;\;b_{2}}^{a_{2}}A_{b_{1}}^{\;\;b_{2}}A_{a_{0}}^{\;\;b_{1}}\gamma
_{a_{2}}.  \label{35}
\end{equation}
Evaluating the Poisson brackets among the constraint functions $\left(
\gamma _{a_{0}},\gamma _{a_{2}}\right) $ with the help of (\ref{34}--\ref{35}%
), we find that they weakly vanish on the surface $\gamma _{a_{0}}\approx 0$%
, $\gamma _{a_{2}}\approx 0$, hence (\ref{17}) and (\ref{27}) are
first-class. This completes the case $L=2$.

\subsection{Generalization to an arbitrary $L$}

Now, we are in the position to generalize the above discussion to an
arbitrary $L$. Acting like in the previous cases, we enlarge the phase-space
with the bosonic canonical pairs $z^{A_{2k+1}}=\left( y^{a_{2k+1}},\pi
_{a_{2k+1}}\right) _{k=0,\cdots ,a}$ and construct an irreducible
Koszul-Tate complex based on the definitions 
\begin{equation}
\delta z^{A}=0,\;\delta z^{A_{2k+1}}=0,\;k=0,\cdots ,a,  \label{36}
\end{equation}
\begin{equation}
\delta {\cal P}_{a_{0}}=-\gamma _{a_{0}},  \label{37}
\end{equation}
\begin{equation}
\delta {\cal P}_{a_{2k}}=-\gamma _{a_{2k}},\;k=1,\cdots ,b,  \label{38}
\end{equation}
which emphasize the irreducible constraints 
\begin{equation}
\gamma _{a_{0}}\equiv G_{a_{0}}+A_{a_{0}}^{\;\;a_{1}}\pi _{a_{1}}\approx 0,
\label{39}
\end{equation}
\begin{equation}
\gamma _{a_{2k}}\equiv Z_{\;\;a_{2k}}^{a_{2k-1}}\pi
_{a_{2k-1}}+A_{a_{2k}}^{\;\;a_{2k+1}}\pi _{a_{2k+1}}\approx 0,\;k=1,\cdots
,b.  \label{40}
\end{equation}
The antighosts $\left( {\cal P}_{a_{2k}}\right) _{k=0,\cdots ,b}$ are all
fermionic with antighost number one, while the notations $a$ and $b$ signify 
\begin{equation}
a=\left\{ 
\begin{array}{c}
\frac{L-1}{2},\;{\rm for}\;L\;{\rm odd}, \\ 
\frac{L}{2}-1,\;{\rm for}\;L\;{\rm even},
\end{array}
\right. \;\;b=\left\{ 
\begin{array}{c}
\frac{L-1}{2},\;{\rm for}\;L\;{\rm odd}, \\ 
\frac{L}{2},\;{\rm for}\;L\;{\rm even}.
\end{array}
\right.  \label{41}
\end{equation}
In order to avoid confusion, we use the conventions $f^{a_{k}}=0$ if $k<0$
or $k>L$. The matrices of the type $A_{a_{k}}^{\;\;a_{k+1}}$ implied in (\ref
{39}--\ref{40}) may involve at most the original variables $z^{A}$ and are
taken to fulfill the relations 
\begin{equation}
rank\left( D_{\;\;b_{k}}^{a_{k}}\right) \approx \sum\limits_{i=k}^{L}\left(
-\right) ^{k+i}M_{i},\;k=1,\cdots ,L-1,  \label{42}
\end{equation}
\begin{equation}
rank\left( D_{\;\;b_{L}}^{a_{L}}\right) =M_{L},  \label{43}
\end{equation}
where $D_{\;\;b_{k}}^{a_{k}}=Z_{\;\;b_{k}}^{a_{k-1}}A_{a_{k-1}}^{\;\;a_{k}}$%
. We remark that the choice of the functions $A_{a_{k-1}}^{\;\;a_{k}}$ is
not unique. Moreover, for a definite choice of these functions, the
equations (\ref{42}--\ref{43}) are unaffected if we modify $%
A_{a_{k-1}}^{\;\;a_{k}}$ as 
\begin{equation}
A_{a_{k-1}}^{\;\;a_{k}}\rightarrow A_{a_{k-1}}^{\;\;a_{k}}+\mu
_{\;\;b_{k-2}}^{a_{k}}Z_{\;\;a_{k-1}}^{b_{k-2}},  \label{45a}
\end{equation}
hence these functions carry some ambiguities.

Acting like in the cases $L=1,2$, after some computation we find the
relations 
\begin{equation}
\pi _{a_{2k+1}}=m_{a_{2k+1}}^{a_{2j}}\gamma _{a_{2j}},\;k=0,\cdots ,a,
\label{44}
\end{equation}
\begin{equation}
G_{a_{0}}=m_{a_{0}}^{a_{2j}}\gamma _{a_{2j}},  \label{45}
\end{equation}
for some functions $m_{a_{2k+1}}^{a_{2j}}$ and $m_{a_{0}}^{a_{2j}}$.
Computing the Poisson brackets among the constraint functions in (\ref{39}--%
\ref{40}), we find that they weakly vanish on the surface (\ref{39}--\ref{40}%
), hence they form a first-class set. This ends the general construction of
an irreducible Koszul-Tate complex associated with the original on-shell $L$%
-stage reducible Hamiltonian system.

\section{Irreducible BRST symmetry for on-shell reducible Hamiltonian systems
}

\subsection{Derivation of the irreducible BRST symmetry}

The first step in deriving an irreducible BRST symmetry for the investigated
on-shell $L$-stage reducible Hamiltonian system has been implemented by
constructing an irreducible Koszul-Tate complex based on the irreducible
first-class constraints (\ref{39}--\ref{40}). The BRST symmetry
corresponding to this irreducible first-class constraint set can be
decomposed like 
\begin{equation}
s=\delta +D+\cdots ,  \label{46}
\end{equation}
where $\delta $ is the irreducible Koszul-Tate differential generated in the
previous section, $D$ represents the longitudinal exterior derivative along
the gauge orbits, and the other pieces, generically denoted by ``$\cdots $%
'', ensure the nilpotency of $s$. The ghost spectrum of the longitudinal
exterior complex includes only the fermionic ghosts $\eta ^{\Delta }\equiv
\left( \eta ^{a_{2k}}\right) _{k=0,\cdots ,b}$ with pure ghost number one ($%
pgh\left( \eta ^{a_{2k}}\right) =1$), respectively associated with the
irreducible first-class constraints (\ref{39}--\ref{40}), to be redenoted by 
$\gamma _{\Delta }\equiv \left( \gamma _{a_{2k}}\right) _{k=0,\cdots ,b}$.
The standard definitions of $D$ acting on the generators from the
longitudinal exterior complex read as 
\begin{equation}
DF=\left[ F,\gamma _{\Delta }\right] \eta ^{\Delta },  \label{47}
\end{equation}
\begin{equation}
D\eta ^{\Delta }=\frac{1}{2}C_{\Delta ^{\prime }\Delta ^{\prime \prime
}}^{\Delta }\eta ^{\Delta ^{\prime }}\eta ^{\Delta ^{\prime \prime }},
\label{48}
\end{equation}
where $F$ can be any function of the variables $z^{A}$, $\left(
z^{A_{2k+1}}\right) _{k=0,\cdots ,a}$, and $C_{\Delta ^{\prime }\Delta
^{\prime \prime }}^{\Delta }$ stand for the first-order structure functions
corresponding to the first-class constraint functions 
\begin{equation}
\left[ \gamma _{\Delta },\gamma _{\Delta ^{\prime }}\right] =C_{\Delta
\Delta ^{\prime }}^{\Delta ^{\prime \prime }}\gamma _{\Delta ^{\prime \prime
}}.  \label{49}
\end{equation}
The action of $\delta $ can be extended to the ghosts through 
\begin{equation}
\delta \eta ^{\Delta }=0,  \label{50}
\end{equation}
with $antigh\left( \eta ^{\Delta }\right) =0$, such that both the nilpotency
and acyclicity of the irreducible Koszul-Tate differential are maintained,
while $D$ can be appropriately extended to the antighosts in such a way to
become a differential modulo $\delta $ on account of the first-class
behaviour of the irreducible constraints. On these grounds, the homological
perturbation theory \cite{13}--\cite{16} guarantees the existence of a
nilpotent BRST symmetry $s$ of the form (\ref{46}) associated with the
irreducible first-class constraints (\ref{39}--\ref{40}).

\subsection{Link with the standard reducible BRST symmetry}

In the sequel we establish the correlation between the standard Hamiltonian
BRST symmetry corresponding to the starting reducible first-class system and
that associated with the irreducible one, investigated in the above
subsection. In this light we show that the physical observables of the two
theories coincide. Let $F$ be an observable of the irreducible system.
Consequently, it satisfies the equations 
\begin{equation}
\left[ F,\gamma _{a_{2k}}\right] \approx 0,\;k=0,\cdots ,b,  \label{55}
\end{equation}
where the weak equality refers to the surface (\ref{39}--\ref{40}). Using
the relations (\ref{44}--\ref{45}), we then find that $F$ also fulfills the
equations 
\begin{equation}
\left[ F,G_{a_{0}}\right] =\left[ F,m_{a_{0}}^{a_{2j}}\right] \gamma
_{a_{2j}}+\left[ F,\gamma _{a_{2j}}\right] m_{a_{0}}^{a_{2j}}\approx 0,
\label{55a}
\end{equation}
\begin{equation}
\left[ F,\pi _{a_{2k+1}}\right] =\left[ F,m_{a_{2k+1}}^{a_{2j}}\right]
\gamma _{a_{2j}}+\left[ F,\gamma _{a_{2j}}\right]
m_{a_{2k+1}}^{a_{2j}}\approx 0,\;k=0,\cdots ,a,  \label{55b}
\end{equation}
on this surface. So, every observable of the irreducible theory should
verify the equations (\ref{55a}--\ref{55b}) on the surface (\ref{39}--\ref
{40}). Now, we observe that the first-class surface described by the
relations (\ref{39}--\ref{40}) is equivalent with that described by the
equations 
\begin{equation}
G_{a_{0}}\approx 0,\;\pi _{a_{2k+1}}\approx 0,\;k=0,\cdots ,a.  \label{55c}
\end{equation}
Indeed, it is clear that when (\ref{55c}) hold, (\ref{39}--\ref{40}) hold,
too. The converse results from (\ref{44}--\ref{45}), which show that if (\ref
{39}--\ref{40}) hold, then (\ref{55c}) also hold. This proves the
equivalence between the first-class surfaces (\ref{39}--\ref{40}) and (\ref
{55c}). As a consequence, we have that every observable of the irreducible
theory, which we found that verifies the equations (\ref{55a}--\ref{55b}) on
the surface (\ref{39}--\ref{40}), will check these equations also on the
surface (\ref{55c}). This means that every observable of the irreducible
system is an observable of the theory based on the first-class constraints (%
\ref{55c}). Conversely, if $F$ represents a physical observable of the
system underlying the constraints (\ref{55c}), then it should check the
equations 
\begin{equation}
\left[ F,G_{a_{0}}\right] \approx 0,\;\left[ F,\pi _{a_{2k+1}}\right]
\approx 0,\;k=0,\cdots ,a,  \label{55d}
\end{equation}
on the surface (\ref{55c}). Then, it follows that $F$ satisfies the
relations 
\begin{equation}
\left[ F,\gamma _{a_{0}}\right] =\left[ F,G_{a_{0}}\right] +\left[
F,A_{a_{0}}^{\;\;a_{1}}\right] \pi _{a_{1}}+\left[ F,\pi _{a_{1}}\right]
A_{a_{0}}^{\;\;a_{1}}\approx 0,  \label{55e}
\end{equation}
\begin{eqnarray}  \label{55f}
& &\left[ F,\gamma _{a_{2k}}\right] =\left[
F,Z_{\;\;a_{2k}}^{a_{2k-1}}\right] \pi _{a_{2k-1}}+\left[ F,\pi
_{a_{2k-1}}\right] Z_{\;\;a_{2k}}^{a_{2k-1}}+  \nonumber \\
& &\left[ F,A_{a_{2k}}^{\;\;a_{2k+1}}\right] \pi _{a_{2k+1}}+\left[ F,\pi
_{a_{2k+1}}\right] A_{a_{2k}}^{\;\;a_{2k+1}}\approx 0,\;k=1,\cdots ,b,
\end{eqnarray}
on the same surface. Recalling once again the equivalence between this
surface and the one expressed by (\ref{39}--\ref{40}), we find that $F$ will
verify the equations (\ref{55e}--\ref{55f}) also on the surface (\ref{39}--%
\ref{40}), being therefore an observable of the irreducible system. From the
above discussion we conclude that the physical observables of the
irreducible theory coincide with those associated with the system subject to
the first-class constraints (\ref{55c}). Next, we show that the physical
observables of the system possessing the constraints (\ref{55c}) and the
ones corresponding to the original reducible theory coincide. In this light,
we remark that the surface (\ref{55c}) can be inferred in a trivial manner
from (\ref{1}) by adding the canonical pairs $\left( y^{a_{2k+1}},\pi
_{a_{2k+1}}\right) _{k=0,\cdots ,a}$ and requiring that their momenta
vanish. Thus, the observables of the original redundant theory are
unaffected by the introduction of the new canonical pairs. In fact, the
difference between an observable $F$ of the system subject to the
constraints (\ref{55c}) and one of the original theory, $\bar{F}$, is of the
type $F-\bar{F}=\sum_{k=0}^{a}f^{a_{2k+1}}\pi _{a_{2k+1}}$. As any two
observables that differ through a combination of first-class constraint
functions can be identified, we find that the physical observables of the
initial theory coincide with those of the system described by the
constraints (\ref{55c}). So far, we proved that the observables of the
system with the constraints (\ref{55c}) coincide on the one hand with those
of the irreducible theory, and, on the other hand, with those of the
original reducible one. In conclusion, the physical observables associated
with the irreducible system also coincide with those of the starting
on-shell reducible first-class theory. In turn, this result will have a
strong impact at the level of the BRST analysis.

In the above we have shown that starting with an arbitrary on-shell
reducible first-class Hamiltonian system displaying the standard Hamiltonian
BRST symmetry $s_{R}$ we can construct a corresponding irreducible
first-class theory, whose BRST symmetry $s$ complies with the basic
requirements of the Hamiltonian BRST formalism. The previous result on the
physical observables induces that the zeroth order cohomological groups of
the corresponding BRST symmetries are isomorphic 
\begin{equation}
H^{0}\left( s_{R}\right) \simeq H^{0}\left( s\right) .  \label{56}
\end{equation}
In addition, both symmetries are nilpotent 
\begin{equation}
s_{R}^{2}=0=s^{2}.  \label{57}
\end{equation}
Then, from the point of view of the fundamental equations of the BRST
formalism, namely, the nilpotency of the BRST operator and the isomorphism
between the zeroth order cohomological group of the BRST differential and
the algebra of physical observables, it follows that it is permissible to
replace the Hamiltonian BRST symmetry of the original on-shell $L$-stage
reducible system with that of the irreducible theory. Thus, we can
substitute the path integral of the reducible system in the Hamiltonian BRST
approach by that of the irreducible theory.

However, it would be convenient to infer a covariant path integral with
respect to the irreducible system. The present phase-space coordinates may
not ensure the covariance. For instance, if we analyze the gauge
transformations of the extended action of the irreducible system, we remark
that those corresponding to the Lagrange multipliers of the constraint
functions $\gamma _{a_{0}}$ will not involve the term $-Z_{\;%
\;a_{1}}^{a_{0}}\epsilon ^{a_{1}}$, which is present within the reducible
context with respect to the constraint functions $G_{a_{0}}$. For all known
models, the presence of this term is essential in arriving at some covariant
gauge transformations at the Lagrangian level. For this reason it is
necessary to gain such a term also within the irreducible setting. Moreover,
it is possible that some of the newly added canonical variables lack
covariant Lagrangian gauge transformations. This signalizes that we need to
add more phase-space variables to be constrained in an appropriate manner.
In view of this, we introduce the additional bosonic canonical pairs 
\begin{equation}
\left( y^{(1)a_{2k+1}},\pi _{a_{2k+1}}^{(1)}\right) ,\;\left(
y^{(2)a_{2k+1}},\pi _{a_{2k+1}}^{(2)}\right) ,\;k=0,\cdots ,a,  \label{57aa}
\end{equation}
\begin{equation}
\left( y^{a_{2k}},\pi _{a_{2k}}\right) ,\;k=1,\cdots ,b,  \label{57ab}
\end{equation}
subject to some constraints of the type 
\begin{equation}
\gamma _{a_{2k+1}}^{(1)}\equiv \pi _{a_{2k+1}}-\pi _{a_{2k+1}}^{(1)}\approx
0,\;\gamma _{a_{2k+1}}^{(2)}\equiv \pi _{a_{2k+1}}^{(2)}\approx
0,\;k=0,\cdots ,a,  \label{57a}
\end{equation}
\begin{equation}
\gamma _{a_{2k}}\equiv \pi _{a_{2k}}\approx 0,\;k=1,\cdots ,b.  \label{57abc}
\end{equation}
In this manner we do not affect in any way the properties of the irreducible
theory as (\ref{57a}--\ref{57abc}) still form together with (\ref{39}--\ref
{40}) an irreducible first-class set. The newly added constraints implies
the introduction of some supplementary ghosts and antighosts and the
extension of the action of the BRST operator on them in the usual manner.
Then, there exists a consistent Hamiltonian BRST symmetry with respect to
the new irreducible theory, described by the constraints (\ref{39}--\ref{40}%
) and (\ref{57a}--\ref{57abc}). Now, if we choose the first-class
Hamiltonian with respect to the above first-class constraints in an adequate
manner, we can in principle generate a gauge algebra that leads to some
covariant Lagrangian gauge transformations. From (\ref{44}) it results that
the former set of constraints in (\ref{57a}) reduces to $\pi
_{a_{2k+1}}^{(1)}\approx 0$. Thus, we observe that the surface (\ref{39}--%
\ref{40}), (\ref{57a}--\ref{57abc}) results in a trivial way from (\ref{39}--%
\ref{40}) by adding the canonical variables (\ref{57aa}--\ref{57ab}) and
demanding that their momenta vanish. Then, the difference between an
observable $F$ of the new irreducible theory and one of the previous
irreducible system, $\bar{F}$, is of the type $F-\bar{F}%
=\sum_{k=0}^{a}f^{a_{2k+1}}\pi
_{a_{2k+1}}^{(1)}+\sum_{k=0}^{a}g^{a_{2k+1}}\pi
_{a_{2k+1}}^{(2)}+\sum_{k=1}^{b}h^{a_{2k}}\pi _{a_{2k}}$, hence $F$ and $%
\bar{F}$ can be identified. Therefore, the physical observables
corresponding to the two irreducible systems coincide, such that the
supplementary constraints (\ref{57a}--\ref{57abc}) do not afflict the
previously established equivalence with the physical observables of the
original redundant theory. In consequence, we can replace the Hamiltonian
BRST symmetry of the original reducible system with that of the latter
irreducible theory, and similarly with regard to the associated path
integrals.

With all these elements at hand, the Hamiltonian BRST quantization of the
irreducible theory goes along the standard lines. Defining a canonical
action for $s$ in the usual way as $s\bullet =\left[ \bullet ,\Omega \right] 
$, with $\Omega $ the canonical generator (the BRST charge), the nilpotency
of $s$ implies that $\Omega $ should satisfy the equation 
\begin{equation}
\left[ \Omega ,\Omega \right] =0.  \label{57xa}
\end{equation}
The existence of the solution to the equation (\ref{57xa}) is then
guaranteed by the acyclicity of the irreducible Koszul-Tate differential at
positive antighost numbers. Once constructed the BRST charge, we pass to the
construction of the BRST-invariant Hamiltonian, $H_{B}=H^{\prime }+{\rm %
``more"}$, that satisfies $\left[ H_{B},\Omega \right] =0$, where $H^{\prime
}$ stands for the first-class Hamiltonian with respect to the constraints (%
\ref{39}--\ref{40}) and (\ref{57a}--\ref{57abc}). In order to fix the gauge,
we have to choose a gauge-fixing fermion $K$ that implements some
irreducible gauge conditions. The possibility to construct some irreducible
gauge conditions is facilitated by the introduction of the pairs $\left(
y,\pi \right) $, which, at this level, play the same role like the auxiliary
variables in the non-minimal approach. Actually, these variables are more
relevant than the corresponding non-minimal ones appearing in the
gauge-fixing process from the reducible case because they prevent the
appearance of the reducibility (via the irreducible first-class
constraints), while the non-minimal coordinates in the reducible situation
are mainly an effect of the redundancy. The gauge-fixed Hamiltonian $%
H_{K}=H_{B}+\left[ K,\Omega \right] $ will then produce a correct
gauge-fixed action $S_{K}$ with respect to the irreducible theory. In this
way, we showed how a reducible first-class Hamiltonian system can be
approach along an irreducible BRST procedure, hence without using either
ghosts for ghosts or their antighosts.

Finally, a word of caution. The appearance of the inverse of the matrices $%
D_{\;\;b_{k}}^{a_{k}}$ in various formulas implies that our approach may
have problems with locality. On the other hand, locality is required for all
concrete applications in field theory. However, taking into account the
ambiguities present in the choice of the reducibility functions \cite{6} and
of the functions $A_{a_{k}}^{\;\;a_{k+1}}$ (see (\ref{45a})), it might be
possible to obtain a local formulation. This completes our treatment.

\section{Example}

In this section we illustrate the general formalism exposed in the above in
the case of a second-stage reducible model. We start with the Lagrangian
action 
\begin{equation}
S_{0}^{L}\left[ A^{\mu \nu \rho }\right] =\int d^{7}x\left( -\frac{1}{48}%
F_{\mu \nu \rho \lambda }F^{\mu \nu \rho \lambda }+\xi \varepsilon _{\mu \nu
\rho \lambda \alpha \beta \gamma }F^{\mu \nu \rho \lambda }A^{\alpha \beta
\gamma }\right) ,  \label{58}
\end{equation}
where 
\begin{equation}
F_{\mu \nu \rho \lambda }=\partial _{\mu }A_{\nu \rho \lambda }-\partial
_{\nu }A_{\mu \rho \lambda }+\partial _{\rho }A_{\lambda \mu \nu }-\partial
_{\lambda }A_{\rho \mu \nu }\equiv \partial _{\left[ \mu \right. }A_{\left.
\nu \rho \lambda \right] },  \label{59}
\end{equation}
$\varepsilon _{\mu \nu \rho \lambda \alpha \beta \gamma }$ represents the
completely antisymmetric seven-dimensional symbol and $\xi $ is a constant.
From the canonical analysis of action (\ref{58}) we deduce the first-class
constraints 
\begin{equation}
\gamma _{ij}^{(1)}\equiv \pi _{0ij}\approx 0,  \label{60}
\end{equation}
\begin{equation}
G_{ij}^{(2)}\equiv -3\left( \partial ^{k}\pi _{kij}+\xi \varepsilon
_{0ijklmn}F^{klmn}\right) \approx 0,  \label{61}
\end{equation}
and the first-class Hamiltonian 
\begin{equation}
H=\int d^{6}x\left( \frac{1}{48}F_{ijkl}^{2}-3\left( \pi _{ijk}-4\xi
\varepsilon _{0ijklmn}A^{lmn}\right) ^{2}+A^{0ij}G_{ij}^{(2)}\right) ,
\label{62}
\end{equation}
where $\pi _{\mu \nu \rho }$ stand for the canonical momenta conjugated with 
$A^{\mu \nu \rho }$. The notation $F_{ijkl}^{2}$ signifies $F_{ijkl}F^{ijkl}$%
, and similarly for the other square. The constraint functions in (\ref{61})
are second-stage reducible, the first-, respectively, second-stage
reducibility relations being expressed by 
\begin{equation}
Z_{\;\;k}^{ij}G_{ij}^{(2)}=0,\;Z^{k}Z_{\;\;k}^{ij}=0,  \label{63}
\end{equation}
where the reducibility functions have the form 
\begin{equation}
Z_{\;\;a_{1}}^{a_{0}}\rightarrow Z_{\;\;k}^{ij}=\partial ^{\left[ i\right.
}\delta _{\;\;k}^{\left. j\right] },\;Z_{\;\;a_{2}}^{a_{1}}\rightarrow
Z^{k}=\partial ^{k}.  \label{64}
\end{equation}

Acting like in section 2, we enlarge the original phase-space with the
bosonic canonical pairs $\left( y^{a_{1}},\pi _{a_{1}}\right) \equiv \left(
H^{i},\pi _{i}\right) $ and construct an irreducible first-class constraint
set corresponding to (\ref{61}) of the type 
\begin{equation}
\gamma _{a_{0}}\rightarrow \gamma _{ij}^{(2)}\equiv -3\left( \partial
^{k}\pi _{kij}+\xi \varepsilon _{0ijklmn}F^{klmn}\right) -\partial _{\left[
i\right. }\pi _{\left. j\right] }\approx 0,  \label{65}
\end{equation}
\begin{equation}
\gamma _{a_{2}}\rightarrow \gamma ^{(2)}\equiv -\partial ^{i}\pi _{i}\approx
0.  \label{66}
\end{equation}
In inferring the above irreducible constraints we took the functions of the
type $A_{a_{k}}^{\;\;a_{k+1}}$ like 
\begin{equation}
A_{a_{0}}^{\;\;a_{1}}\rightarrow A_{ij}^{\;\;k}=\partial _{\left[ i\right.
}\delta _{\left. j\right] }^{\;\;k},\;A_{a_{1}}^{\;\;a_{2}}\rightarrow
A_{k}=\partial _{k},  \label{66a}
\end{equation}
such that they comply with the requirements from the general theory. As we
mentioned in section 3, in order to generate a covariant gauge-fixed action
as a result of the Hamiltonian BRST quantization of the irreducible model it
is still necessary to enlarge the phase-space with some canonical pairs
subject to additional constraints such that the overall constraint set is
first-class and irreducible. In this light, we introduce the canonical pairs 
$\left( H^{0},\pi _{0}\right) $, $\left( H^{(1)i},\pi _{i}^{(1)}\right) $, $%
\left( H^{(2)i},\pi _{i}^{(2)}\right) $ constrained like 
\begin{equation}
\gamma ^{(1)}\equiv \pi _{0}\approx 0,\;\gamma _{i}^{(1)}\equiv \pi _{i}-\pi
_{i}^{(1)}\approx 0,\;\gamma _{i}^{(2)}\equiv -2\pi _{i}^{(2)}\approx 0.
\label{67}
\end{equation}
Obviously, the constraint set (\ref{60}), (\ref{65}--\ref{66}), (\ref{67})
is irreducible and first-class. We take the first-class Hamiltonian with
respect to this set under the form 
\begin{eqnarray}  \label{68}
& &H^{\prime }=\int d^{6}x\left( \frac{1}{48}F_{ijkl}^{2}-3\left( \pi
_{ijk}- 4\xi \varepsilon _{0ijklmn}A^{lmn}\right) ^{2}+ A^{0ij}\gamma
_{ij}^{(2)}+ \right.  \nonumber \\
& &\left. H^{0}\gamma ^{(2)}+H^{i}\pi _{i}^{(2)}+H^{(2)i}\left( \partial
^{j}\gamma _{ji}^{(2)}+\partial _{i}\gamma ^{(2)}\right) \right) ,
\end{eqnarray}
such that the irreducible gauge algebra will lead to a covariant path
integral.

Next, we approach the Hamiltonian BRST quantization of the irreducible
model. In view of this, we add the minimal fermionic canonical pairs
ghost-antighost $\left( \eta ^{(\Delta )ij},{\cal P}_{ij}^{(\Delta )}\right) 
$, $\left( \eta ^{(\Delta )i},{\cal P}_{i}^{(\Delta )}\right) $, $\left(
\eta ^{(\Delta )},{\cal P}^{(\Delta )}\right) $, with $\Delta =1,2$,
associated with the corresponding constraints in (\ref{60}), (\ref{65}--\ref
{66}) and (\ref{67}). All the ghosts possess ghost number one, while their
antighosts display ghost number minus one. Moreover, we take a non-minimal
sector organized as $\left( P_{b}^{ij},b_{ij}\right) $, $\left(
P_{b},b\right) $, $\left( P_{b^{1}}^{ij},b_{ij}^{1}\right) $, $\left(
P_{b^{1}},b^{1}\right) $, $\left( P_{\bar{\eta}}^{ij},\bar{\eta}_{ij}\right) 
$, $\left( P_{\bar{\eta}},\bar{\eta}\right) $, $\left( P_{\bar{\eta}%
^{1}}^{ij},\bar{\eta}_{ij}^{1}\right) $, $\left( P_{\bar{\eta}^{1}},\bar{\eta%
}^{1}\right) $. The first four sets of non-minimal variables are bosonic
with ghost number zero, while the others are fermionic, with the $P_{\bar{%
\eta}}$'s and $\bar{\eta}$'s of ghost number one, respectively, minus one.
Consequently, the non-minimal BRST charge reads as 
\begin{eqnarray}  \label{69}
& &\Omega =\int d^{6}x\left( \sum\limits_{\Delta =1}^{2}\left( \gamma
_{ij}^{(\Delta )}\eta ^{(\Delta )ij}+ \gamma _{i}^{(\Delta )}\eta ^{(\Delta
)i}+ \gamma ^{(\Delta )}\eta ^{(\Delta )}\right) +\right.  \nonumber \\
& &\left. P_{\bar{\eta}}^{ij}b_{ij}+P_{\bar{\eta}}b+ P_{\bar{\eta}%
^{1}}^{ij}b_{ij}^{1}+ P_{\bar{\eta}^{1}}b^{1}\right) .
\end{eqnarray}
The BRST-invariant Hamiltonian corresponding to (\ref{68}) is given by 
\begin{eqnarray}  \label{70}
& &H_{B}=\int d^{6}x\left( \eta ^{(1)ij}{\cal P}_{ij}^{(2)}+ \eta ^{(1)}%
{\cal P}^{(2)}-\frac{1}{2}\eta ^{(1)i} {\cal P}_{i}^{(2)}+\frac{1}{2}\eta
^{(2)ij} \partial _{\left[ i\right. }{\cal P}_{\left. j\right] }^{(2)}+
\right.  \nonumber \\
& &\left. \frac{1}{2}\eta ^{(2)}\partial ^{i}{\cal P}_{i}^{(2)}- 2\eta
^{(2)i}\left( \partial ^{j}{\cal P}_{ji}^{(2)}+ \partial _{i}{\cal P}%
^{(2)}\right) \right) +H^{\prime }.
\end{eqnarray}
In order to determine the gauge-fixed action, we work with the gauge-fixing
fermion 
\begin{eqnarray}  \label{71}
& &K=\int d^{6}x\left( {\cal P}_{ij}^{(1)}\left( \partial _{k} A^{kij}+\frac{%
1}{2}\partial ^{\left[ i\right. } H^{(1)\left. j\right] }\right) +{\cal P}%
^{(1)}\partial _{i}H^{(1)i}+ \right.  \nonumber \\
& &{\cal P}_{i}^{(1)}\left( 2\partial _{j}A^{ji0}+ \partial ^{i}H^{0}\right)
+P_{b^{1}}^{ij}\left( {\cal P}_{ij}^{(1)}-\bar{\eta}_{ij}+ \stackrel{.}{\bar{%
\eta}}_{ij}^{1}\right) +  \nonumber \\
& &\left. P_{b^{1}}\left( {\cal P}^{(1)}-\bar{\eta}+ \stackrel{.}{\bar{\eta}}%
^{1}\right) + P_{b}^{ij}\left( \bar{\eta}_{ij}^{1}+ \stackrel{.}{\bar{\eta}}%
_{ij}\right) + P_{b}\left( \bar{\eta}^{1}+ \stackrel{.}{\bar{\eta}}\right)
\right) .
\end{eqnarray}
It is clear that the above gauge-fixing fermion implements some irreducible
canonical gauge conditions. Introducing the expression of the gauge-fixed
Hamiltonian $H_{K}=H_{B}+\left[ K,\Omega \right] $ in the gauge-fixed action
and eliminating some auxiliary variables on their equations of motion, we
finally find the path integral 
\begin{equation}
Z_{K}=\int {\cal D}A^{\mu \nu \rho }{\cal D}H^{(1)\mu }{\cal D}b_{\mu \nu }%
{\cal D}b{\cal D}\eta ^{(2)\mu \nu }{\cal D}\eta ^{(2)}{\cal D}\bar{\eta}%
_{\mu \nu }{\cal D}\bar{\eta}\exp \left( iS_{K}\right) ,  \label{72}
\end{equation}
where 
\begin{eqnarray}  \label{73}
& &S_{K}=S_{0}^{L}\left[ A^{\mu \nu \rho }\right] + \int d^{7}x\left( b_{\mu
\nu }\left( \partial _{\rho } A^{\rho \mu \nu }+\frac{1}{2}\partial ^{\left[
\mu \right. } H^{(1)\left. \nu \right] }\right) +\right.  \nonumber \\
& &\left. b\partial _{\mu }H^{(1)\mu }-\bar{\eta}_{\mu \nu } \Box \eta
^{(2)\mu \nu }-\bar{\eta}\Box \eta ^{(2)}\right) .
\end{eqnarray}
In deriving (\ref{73}) we performed the identifications 
\begin{equation}
H^{(1)\mu }=\left( H^{0},H^{(1)i}\right) ,\;b_{\mu \nu }=\left( \pi
_{i}^{(1)},b_{ij}\right) ,  \label{74}
\end{equation}
\begin{equation}
\bar{\eta}_{\mu \nu }=\left( -{\cal P}_{i}^{(1)},\bar{\eta}_{ij}\right)
,\eta ^{(2)\mu \nu }=\left( \eta ^{(2)i},\eta ^{(2)ij}\right) ,  \label{75}
\end{equation}
and used the symbol $\Box =\partial _{\rho }\partial ^{\rho }$. It is clear
that the gauge-fixed action (\ref{73}) is covariant and displays no residual
gauge invariances although we have not used any ghosts for ghosts. It is
precisely the introduction of the supplementary canonical pairs constrained
accordingly to (\ref{67}) and the choice (\ref{68}) of the first-class
Hamiltonian that generates a Hamiltonian gauge algebra implementing the
covariance. Actually, the gauge-fixed action (\ref{73}) can be alternatively
inferred in the framework of the antifield BRST treatment if we start with
the action 
\begin{equation}
S_{0}^{L}\left[ A^{\mu \nu \rho },H^{(1)\mu }\right] =S_{0}^{L}\left[ A^{\mu
\nu \rho }\right] ,  \label{75a}
\end{equation}
subject to the irreducible gauge transformations 
\begin{equation}
\delta _{\epsilon }A^{\mu \nu \rho }=\partial ^{\left[ \mu \right. }\epsilon
^{\left. \nu \rho \right] },\;\delta _{\epsilon }H^{(1)\mu }=2\partial _{\nu
}\epsilon ^{\nu \mu }+\partial ^{\mu }\epsilon ,  \label{75b}
\end{equation}
and employ an adequate non-minimal sector and gauge-fixing fermion. In fact,
the irreducible and covariant gauge transformations (\ref{75b}) are the
Lagrangian result of our irreducible Hamiltonian procedure inferred via the
extended and total actions and their gauge transformations.

In the end, let us briefly compare the above results with those derived in
the standard reducible BRST approach. The gauge-fixed action in the usual
reducible BRST framework can be brought to the form 
\begin{eqnarray}  \label{76}
& &S_{\psi }=S_{0}^{L}\left[ A^{\mu \nu \rho }\right] + \int d^{7}x\left( 
{\cal B}_{\mu \nu }\left( \partial _{\rho }A^{\rho \mu \nu }+\frac{1}{2}
\partial ^{\left[ \mu \right. } \varphi ^{\left. \nu \right] }\right)
+\right.  \nonumber \\
& &\left. {\cal B}\partial _{\mu }\varphi ^{\mu }- \bar{C}_{\mu \nu }\Box
C^{\mu \nu }- \bar{C}^{2}\Box \bar{C}^{1}- \bar{C}_{\mu }\Box C^{\mu }-\bar{C%
}\Box C\right) ,
\end{eqnarray}
where $C^{\mu \nu }$ stand for the ghost number one ghosts, $C^{\mu }$
represent the ghost number two ghosts for ghosts, and $C$ is the ghost
number three ghost for ghost for ghost. The remaining variables belong to
the non-minimal sector and have the properties 
\begin{equation}
\epsilon \left( {\cal B}_{\mu \nu }\right) =0,\;gh\left( {\cal B}_{\mu \nu
}\right) =0,\;\epsilon \left( \varphi ^{\mu }\right) =0,\;gh\left( \varphi
^{\mu }\right) =0,  \label{77}
\end{equation}
\begin{equation}
\epsilon \left( {\cal B}\right) =0,\;gh\left( {\cal B}\right) =0,\;\epsilon
\left( \bar{C}_{\mu \nu }\right) =1,\;gh\left( \bar{C}_{\mu \nu }\right) =-1,
\label{78}
\end{equation}
\begin{equation}
\epsilon \left( \bar{C}_{\mu }\right) =0,\;gh\left( \bar{C}_{\mu }\right)
=-2,\;\epsilon \left( \bar{C}\right) =1,\;gh\left( \bar{C}\right) =-3,
\label{79}
\end{equation}
\begin{equation}
\epsilon \left( \bar{C}^{2}\right) =1,\;gh\left( \bar{C}^{2}\right)
=-1,\;\epsilon \left( \bar{C}^{1}\right) =1,\;gh\left( \bar{C}^{1}\right) =1.
\label{80}
\end{equation}
By realizing the identifications 
\begin{equation}
C^{\mu \nu }\leftrightarrow \eta ^{(2)\mu \nu },\;{\cal B}_{\mu \nu
}\leftrightarrow b_{\mu \nu },\;\varphi ^{\mu }\leftrightarrow H^{(1)\mu },
\label{81}
\end{equation}
\begin{equation}
{\cal B}\leftrightarrow b,\;\bar{C}_{\mu \nu }\leftrightarrow \bar{\eta}%
_{\mu \nu },\;\bar{C}^{2}\leftrightarrow \bar{\eta},\;\bar{C}%
^{1}\leftrightarrow \eta ^{(2)},  \label{82}
\end{equation}
among the variables involved with the gauge-fixed actions inferred within
the irreducible and reducible approaches, (\ref{73}), respectively, (\ref{76}%
), the difference between the two gauge-fixed actions is given by 
\begin{equation}
S_{K}-S_{\psi }=\int d^{7}x\left( \bar{C}_{\mu }\Box C^{\mu }+\bar{C}\Box
C\right) .  \label{83}
\end{equation}
We remark that $S_{K}-S_{\psi }$ is proportional to the ghosts for ghosts $%
C^{\mu }$ and the ghost for ghost for ghost $C$, which are some essential
compounds of the reducible BRST quantization. Although identified at the
level of the gauge-fixed actions, the fields $\varphi ^{\mu }$ and $%
H^{(1)\mu }$ play different roles within the two formalisms. More precisely,
the presence of $H^{(1)\mu }$ within the irreducible model prevents the
appearance of the reducibility, while $\varphi ^{\mu }$ is nothing but one
of its effects. In fact, the effect of introducing the fields $H^{(1)\mu }$
is that of forbidding the appearance of zero modes which exist within the
original reducible theory by means of the first term present in the gauge
transformations of $H^{(1)\mu }$ from (\ref{75b}). Indeed, if we take $%
\epsilon ^{\mu \nu }=\partial ^{\left[ \mu \right. }\epsilon ^{\left. \nu
\right] }$ in (\ref{75b}), then $\partial _{\mu }\epsilon ^{\mu \nu }$ is
non-vanishing, such that the entire set of gauge transformations is
irreducible. In consequence, all the quantities linked with zero modes, like
the ghosts with ghost number greater than one or the pyramid-like structured
non-minimal sector, are discarded when passing to the irreducible setting.
This completes the analysis of the investigated model.

\section{Conclusion}

In conclusion, in this paper we have shown how systems with reducible
first-class constraints can be quantized by using an irreducible Hamiltonian
BRST formalism. The key point of our approach is given by the construction
of a Hamiltonian Koszul-Tate complex that emphasizes an irreducible set of
first-class constraints. As the physical observables associated with the
irreducible and reducible versions coincide, the main equations underlining
the Hamiltonian BRST formalism make legitimate the substitution of the BRST
quantization of the reducible theory by that of the irreducible one. The
canonical generator of the irreducible BRST symmetry exists due to the
acyclicity of the irreducible Koszul-Tate differential, while the
gauge-fixing procedure is facilitated by the enlargement of the phase-space
with the canonical pairs of the type $\left( y,\pi \right) $. The general
formalism is exemplified on a second-stage reducible model involving
three-form gauge fields, which is then compared with the results from the
standard reducible BRST approach.

\section*{Acknowledgment}

This work has been supported by a Romanian National Council for Academic
Scientific Research (CNCSIS) grant.

\end{document}